\documentclass[twocolumn,usenatbib,amsmath,amssymb,amsbsy]{mn2e}

\usepackage{natbib}
\usepackage{amsbsy}
\usepackage{amssymb}
\usepackage{amsmath}

\usepackage{graphicx}

\newcommand{\be}{\begin{equation}}
\newcommand{\ee}{\end{equation}}
\newcommand{\bear}{\begin{eqnarray}}
\newcommand{\eear}{\end{eqnarray}}

\newcommand{\xp}{x_{\rm p}}

\newcommand{\rx}{{\rm x}}

\newcommand{\rn}{{\rm n}}
\newcommand{\rp}{{\rm p}}
\newcommand{\re}{{\rm e}}
\newcommand{\rc}{{\rm c}}

\newcommand{\mut}{\tilde{\mu}}
\newcommand{\dmut}{{\delta \mut}}

\newcommand{\epstar}{{\varepsilon_\star}}

\newcommand{\rL}{{\rm L}}

\newcommand{\rP}{{\rm P}}
\newcommand{\rT}{{\rm T}}

\newcommand{\cA}{{\cal A}}

\newcommand{\cI}{{\cal I}}

\begin{document}

\title[MHD equilibrium in multifluid neutron stars]{Hydromagnetic equilibrium in non-barotropic multifluid neutron stars}

\author[Glampedakis, Andersson \& Lander]{K. Glampedakis$^1$, N. Andersson$^2$ \&  S.K. Lander$^{2, 3}$\\
  \\
  $^1$  Theoretical Astrophysics, University of T\"ubingen, Auf der Morgenstelle 10, T\"ubingen D-72076, Germany \\
  $^2$ Mathematical Sciences, University of Southampton, Southampton
  SO17 1BJ, UK \\
  $^3$ Max-Planck-Institut f\"ur Gravitationphysik, Albert-Einstein-Institut, Potsdam D-14476, Germany }

\maketitle

\begin{abstract}

Traditionally, the subject of hydromagnetic equilibrium in neutron stars has been addressed in the context of standard 
magnetohydrodynamics, with matter obeying a barotropic equation of state. In this paper we take a step towards a more realistic 
treatment of the problem by considering neutron stars with interior superfluid components. In this multifluid model 
stratification associated with a varying matter composition (the relative proton to neutron density fraction) enters as a natural ingredient, 
leading to a non-barotropic system. After formulating the hydromagnetic equilibrium of superfluid/superconducting neutron stars 
as a perturbation problem, we focus on the particular case of a three-fluid system consisting of superfluid neutrons and normal
protons and electrons. We determine the equilibrium structure of dipolar magnetic fields with a mixed poloidal-toroidal composition. 
We find that, with respect to barotropic models, stratification has the generic effect of leading to equilibria with a higher 
fraction of magnetic energy stored in the toroidal component. However, even in models with strong stratification the poloidal and 
toroidal components are comparable, with the former contributing the bulk of the magnetic energy.

\end{abstract} 
  
\begin{keywords}
stars: magnetars -- stars: neutron -- stars: magnetic fields
\end{keywords}


\section{Introduction}
\label{sec:intro}

Neutron stars exhibit a rich phenomenology, observed in a variety of channels. The quality of the associated data has improved considerably 
in the last decade, and we are now beginning to make detailed inferences about the complex physics associated with these systems. 
A notable recent ÒsuccessÓ concerns the evidence of superfluidity in the compact remnant in the Cassiopeia A, the youngest observed neutron star 
in the Galaxy \citep{page11,shtern11}. 
Nevertheless, we are quite far from a truly quantitative understanding of these objects. Issues associated with the, sometimes 
extremely strong, magnetic field are particularly vexing. The magnetic field is invoked to explain a range of observations, yet we do not have 
particularly reliable theoretical models. Most notably, despite more than 40 years of observations, the mechanism 
that leads to the observed radio pulses remains a puzzle. A similar problem concerns the observed magnetar giant flares. 
In particular, we do not have a clear picture of the emission mechanism associated with the quasiperiodic oscillations seen in the light curve 
of these events. The observed frequencies seem to fit the spectrum of magneto-elastic oscillations associated with the magnetic field and the 
elastic crust, but how exactly do these mechanical vibrations give rise to the observed X-ray variability? 
Moving inwards, the physics of the neutron star core also remains uncertain. The structure of the interior magnetic field depends crucially 
on the composition and the state of matter at extreme densities. 
It is generally expected that mature neutron stars are sufficiently cold to have superfluid/superconducting cores \citep{baym,supercon}. 
If this is the case, then we need to understand how superconductivity impacts on the long-term magnetic field evolution, for example, 
by expelling the interior field on an astrophysically relevant timescale.
Alternatively, we need to understand how the expected quantised fluxtubes alter the magnetic forces 
and (perhaps) the structure of the global field. These issues depend crucially on the matter composition, and the associated critical 
temperature/density for the transition to superfluidity. Current state-of-the-art magnetic star modeling is, however, not yet at this level. 
The stark reality is that we do 
not even understand the relevance of varying matter composition. Existing models have, almost exclusively, considered barotropic fluid models,
that is, models obeying an equation of state relation $p(\rho)$ between the fluid pressure and density
(for the most recent work see \citet{tomi05,haskell08,ciolfi09,lander09,ciolfi10,lasky11,ciolfi11}; references to earlier work can be found
in \citet{Mestelbook}).
Yet, we know that composition variations are important for neutron star dynamics, and it is generally expected that there may be an impact on the magnetic 
field structure as well. This has been pointed out by \citet{reisenegger09} and, to some extent, stratification has been incorporated in the numerical
models of Braithwaite (see for example \citet{braith06,braith09}) albeit in terms of a radial entropy gradient for an ideal gas equation of state. 

If we want to be able to consider realistic models for mature neutron stars, then we need to be able to construct non-barotropic 
magnetic neutron star models. 
This is the aim of the present work. Working in the framework of perturbation theory, which should be adequate for realistic neutron stars, 
we will develop models for the global magnetic field that takes full account of the multifluid nature of 
superfluid neutron stars and the associated variation in the internal composition. 

In this paper we focus on hydromagnetic models with a neutron superfluid component while `ignoring' proton superconductivity. 
The magnetic field is assumed to have a dipolar structure while for the stellar matter we employ a simple polytropic equation of state and a 
phenomenological expression for the stratification associated with the proton fraction. A companion paper \citep{LAG} extends the modelling to a 
non-perturbative 
framework with a generic magnetic field geometry, a broader class of non-barotropic equations of state and includes a first discussion of the 
challenging issue of hydromagnetic equilibrium in superconducting neutron stars.


\section{The multifluid model}
\label{sec:2fluids}

Our neutron star model is a multifluid system consisting of neutrons, protons and electrons (e.g. \citet{nagc06}).
The system is assumed to be {\em axisymmetric} with the fluids at rest. This last assumption is
reasonable for slowly spinning neutron stars like the magnetars where the magnetic field energy
is much larger than the rotational kinetic energy \citep{DT92}. This work is primarily focused on those 
systems.

In addition, we make the following assumptions: We work in the context of Newtonian gravity 
which is justified given the overall level of precision of our calculation. We also assume 
the presence of neutron superfluidity in the entire stellar volume. Real systems, due to the density dependence of the critical temperatures
for superfluidity/superconductivity, will have distinct super- and normal fluid regions. We ignore the presence of the 
elastic crust. This is an appropriate approximation as long as we focus on equilibrium configurations and the crust is in a relaxed state.

\subsection{Basic formalism}
\label{sec:formalism}

The hydromagnetic equilibrium is governed by two coupled Euler equations (see \citet{supercon} 
for details):
\be
{\bf \nabla} \left ( \mut_\rx + \Phi \right ) = \frac{1}{\rho_\rx} {\bf F}_\rx, \qquad \rx = \{\rn,\rp\}
\label{euler0}
\ee
In terms of the particle chemical potentials $\mu_\rn, \mu_\rp$ and $\mu_\re$ and the mass $m=m_\rn = m_\rp$ we have defined
\be
\mut_\rn = \frac{\mu_\rn}{m}, \qquad \mut_\rp = \frac{1}{m} \left ( \mu_\rp + \mu_\re \right )
\ee
The Euler equations also feature the fluid densities $\rho_\rn, \rho_\rp$, the gravitational potential $\Phi$ 
and the magnetic forces ${\bf F}_\rx $ (to be specified below). The total fluid pressure is given by the thermodynamical 
relation \citep{prix04}
\be
{\bf \nabla} p = \rho_\rn {\bf  \nabla} \mut_\rn + \rho_\rp {\bf  \nabla} \mut_\rp 
\label{press_def}
\ee

Our approach to the problem of hydromagnetic equilibrium is based on treating the magnetised system as a perturbation about 
a non-magnetic and spherically symmetric ``background'' configuration.
That is, we write $\mut_\rx \to \mut_\rx + \dmut_\rx$ etc. and work at leading order with respect to the perturbations.
This approximation implies that the magnetic forces should only appear in the perturbed part of equation (\ref{euler0}).
As an additional simplification we will assume that the star retains its spherical shape even in the presence of the
magnetic field. This is an accurate approximation even for the strongest observed magnetic fields (for example, 
see \citet{haskell08}).

To begin with, it is easy to show that the non-magnetic background configuration is in chemical beta-equilibrium, that is,
$ \mut_\rn = \mut_\rp \equiv \mut$. This follows from eqn.~(\ref{euler0}) after setting ${\bf F}_\rx =0$, and taking the difference of the 
two equations. 
From (\ref{press_def}) it follows that
\be
{\bf \nabla} \delta p = \rho_\rn {\bf \nabla} \dmut_\rn + \rho_\rp {\bf  \nabla} \dmut_\rp 
+ \delta \rho {\bf \nabla} \mut
\ee
The background pressure is then simply
\be
{\bf  \nabla} p = \rho {\bf  \nabla} \mut
\ee
where $\rho = \rho_\rn + \rho_\rp $ is the total density.

In formulating the perturbation equations it is convenient to use the perturbed enthalpy
\be
\delta h = \frac{\delta p}{\rho} = (1-\xp) \dmut_\rn + \xp \dmut_\rp
\ee
where we have introduced the proton fraction
\be
\xp = \frac{\rho_\rp}{\rho}
\ee
We then have
\be
\rho_\rn {\bf  \nabla} \dmut_\rn + \rho_\rp {\bf \nabla} \dmut_\rp = \rho \left (  {\bf \nabla} \delta h 
- \delta \beta {\bf \nabla} \xp \right )
\ee
where $\delta \beta = \dmut_\rp -\dmut_\rn $ represents the departure from chemical equilibrium induced by the perturbations.

The perturbed magnetic equilibrium is then determined by the Euler equations
\be
{\bf \nabla} \left ( \dmut_\rx + \delta\Phi  \right ) = \frac{1}{\rho_\rx} {\bf F}_\rx, \qquad \rx = \{ \rn,\rp \}
\label{euler1}
\ee
Note that the densities $\rho_\rx$ appearing here refer to the {\em non-magnetic} system.
These equations are supplemented by the Poisson equation for the gravitational potential,
\be
\nabla^2 \delta \Phi = 4\pi G \delta \rho
\label{poisson}
\ee

Following a strategy familiar from the two-fluid oscillation problem (e.g. \citet{agh09})
we can combine the equations (\ref{euler1}) and produce an equivalent pair of ``average''  and ``difference'' equations.
These are
\bear
&& {\bf \nabla} ( \delta h + \delta \Phi ) -\delta \beta {\bf \nabla} \xp 
= \frac{1}{\rho} \left ( {\bf F}_\rp + {\bf F}_\rn \right ) 
\label{aveuler}
\\
\nonumber \\
&& {\bf\nabla} \delta \beta  = \frac{1}{\rho_\rp} {\bf F}_\rp - \frac{1}{\rho_\rn} {\bf F}_\rn
\label{diffeuler}
\eear
The first equation is essentially the familiar single-fluid Euler equation. Notice, however, 
that apart from the pressure and gravitational forces this equation contains an additional
force term due to ${\bf\nabla} \xp$. This extra term represents the effect of the 
{\em composition stratification} in the hydromagnetic equilibrium. As discussed by \citet{RG92}
and \citet{rieutord}, in the context of neutron star dynamics, this term plays the role of an effective buoyancy, 
providing the restoring force for the family of $g$-modes in oscillating stratified stars.

This discussion highlights the important fact that a stratified system is {\em non-barotropic}, that is,
the matter is not described by a single parameter equation of state of the form $p=p\,(\rho)$. In the present multifluid system
the chemical potentials depend on both densities, i.e. are functionals of the form $\mut_\rx (\rho_\rn,\rho_\rp)$. As a result, the equation 
of state is of the biparametric form $p=p\,(\rho,\beta)$ or, equivalently, $p=p\,(\rho,\xp)$ \citep{RG92,prix04}.

The difference Euler equation (\ref{diffeuler}) is `unique' to 
multifluid systems. In the present context it describes the departure $\delta \beta$ from chemical equilibrium driven 
by the total magnetic force. This heterogeneous magneto-chemical balance is the main driving agent for
magnetic ambipolar diffusion in neutron stars \citep{GR92,GJS11}. The fluid motion associated with ambipolar
diffusion is ignored in the present model. This is equivalent to neglecting nuclear reactions involving
the particles. Inverting this argument, we would expect particle reactions (which should be taking place 
in real neutron stars) to induce fluid flow (this is clear from the mass continuity equation
associated with each fluid). This flow would also induce frictional forces appearing in the Euler equation (\ref{diffeuler}).
The physics of ambipolar flow in superfluid neutron stars was recently discussed in detail by  \citet{GJS11},
who showed that the hydromagnetic quasi-equilibrium is still well described by (\ref{aveuler}) and (\ref{diffeuler})
despite the presence of ambipolar diffusion. In the terminology of \citet{GJS11} this is a `tranfusion-dominated'
quasi-equilibrium.


\subsection{The magnetic forces}
\label{sec:forces}

The magnetic forces ${\bf F}_\rx$ entering in the multifluid formalism 
are directly linked to the properties of neutron star matter, specifically to proton superconductivity and 
neutron superfluidity. 

Neutron stars are expected to manifest both these properties in the bulk of their interiors provided their temperature lies below 
the critical temperatures $T_{\rm c} \sim 10^9\,\mbox{K}$ for the onset of proton and neutron superfluidity. According to 
standard cooling theory this should happen roughly a year after the star is born \citep{cool}. Theory also suggests that the proton 
superconductivity is of the type II \citep{baym}, which means that the magnetic field penetrates the matter by forming a large
number of quantised magnetic (proton) fluxtubes. 

The magnetohydrodynamics of a multifluid system with a superconducting component 
is known to be significantly different, both qualitatively and quantitatively, from standard magnetohydrodynamics 
\citep{easson77,mendell91,mendell98,supercon}. A fundamental difference is the magnetic force itself.

In the presence of type II proton superconductivity the magnetic force exerted on the proton-electron plasma 
(denoted as ${\bf F}_\rp$ in the previous Section) is no longer given by the familiar Lorentz force. Instead, the superconducting force
originates from the tension (energy per unit length) associated with the proton fluxtube array. 
Its explicit form is \citep{supercon}:
\be
{\bf F}_\rp = -\frac{1}{4\pi} \left [\, {\bf B} \times ( {\bf \nabla} \times {\bf H}_{\rc1} ) +
\rho_\rp {\bf\nabla} \left ( B \frac{\partial H_{\rc1}}{\partial \rho_\rp}   \right )  \right ]
\label{Fpsupercon} 
\ee
where ${\bf B}$ is the smooth-averaged magnetic field and ${\bf H}_{\rc1} \equiv (H_{\rc1} /B) {\bf B} $. 
The critical field $H_{\rc1} $ is directly related to the energy per unit length ${\cal E}_\rp$ of each 
individual proton fluxtube and is given by \citep{tilley,supercon}
\be
H_{\rc1} = \frac{4\pi}{\phi_0} {\cal E}_\rp = h_c \frac{\rho_\rp}{\epstar}
\label{Hc}
 \ee
where $\phi_0 = hc/2e$ is the magnetic flux quantum associated with a fluxtube. The definition of 
$h_c \approx \mbox{constant}$ and of the entrainment parameter $\epstar (\rho_\rn,\rho_\rp)$ can be found in
\citet{supercon}.

Somewhat counter-intuitively, the neutron superfluid also experiences a magnetic force ${\bf F}_\rn$. 
This force originates from the coupling of the neutron fluid to the proton fluxtubes via the entrainment parameter $\epstar$, and is 
given by \citep{supercon}
\be
{\bf F}_\rn = -\frac{\rho_\rn}{4\pi} {\bf \nabla} \left ( B \frac{\partial H_{\rc1}}{\partial \rho_\rn}   \right )
\label{Fnsuperon}
\ee
It is worth noting that, with respect to their magnitude, $F_\rn \sim F_\rp$ which indicates that both forces are equally
important in the equilibrium of superconducting/superfluid neutron stars. 

A second scenario to be considered is the state where there is proton superconductivity without the simultaneous presence 
of neutron superfluidity.
This could be an astrophysically relevant case if the neutron star core were to be relatively hot, with a temperature $T$ somewhere in the
range $ 5 \times 10^8\, \mbox{K} \lesssim T \lesssim  10^9\,\mbox{K}$. This temperature range is identified by the recent cooling 
observations of the neutron star located in Cassiopeia A as the range for the onset of neutron superfluidity 
\citep{page11,shtern11}. In this scenario the magnetic force exerted on the proton fluid is again given by (\ref{Fpsupercon}); 
this time, however, the entrainment effect between neutrons and protons on the mesoscopic scale of individual fluxtubes is absent. 
This leads to $\epstar =1$ in (\ref{Hc}) and the  decoupling of the neutron fluid and the magnetic field, 
i.e. ${\bf F}_\rn = 0$. 

A third possible state of matter is that of a system consisting of superfluid neutrons and {\em normal} (non-superconducting) protons.
This could be realised in neutron stars provided the bulk magnetic field strength
in the stellar interior exceeds the critical threshold $H_{\rm c2} \approx 10^{16}\,\mbox{G}$ 
above which superconductivity is suppressed regardless of temperature \citep{baym}. With superconductivity absent, the magnetic field 
behaves ``classically'' and ${\bf F}_\rp$ can be identified with the usual Lorentz force ${\bf F}_\rL$,
\be
{\bf F}_\rp = {\bf F}_\rL =  \frac{1}{4\pi} ({\bf \nabla} \times {\bf B}) \times {\bf B}  
\ee
At the same time, the neutron fluid is oblivious to the presence of the magnetic field, i.e. ${\bf F}_\rn = 0$.


\section{Equilibrium without superconductivity}

Out of the three possible combinations of neutron superfluidity/proton superconductivity the first is likely to be the most relevant 
for neutron stars (and in particular for magnetars). Based on the present understanding of the involved physics, we expect  the simultaneous presence of 
superfluidity and 
superconductivity. Nevertheless,  we will focus on the third scenario, where superconductivity is absent. 
We have a good reason for doing so: the calculation of the full 
superconducting/superfluid hydromagnetic equilibrium represents a big leap with respect to the existing work on the subject which
has been limited to ordinary\footnote{An exception is the recent work of \citet{akgun08} on toroidal fields in a single-component 
type II superconducting barotropic neutron star model.} single-fluid barotropic neutron star models (e.g. 
\citet{haskell08,ciolfi09,ciolfi10,lander09}). By considering a multifluid system with the proton superconductivity ``switched-off'' we can 
advance our understanding of the role of the stratification in the proton fraction considerably.

The hydromagnetic equilibrium for the chosen multifluid system is described by
\bear
&& {\bf \nabla} ( \delta h + \delta \Phi ) -\delta \beta {\bf \nabla} \xp 
= \frac{1}{\rho} {\bf F}_\rL
\label{avEuler}
\\
\nonumber \\
&& {\bf \nabla} \delta \beta  = \frac{1}{\rho_\rp} {\bf F}_\rL 
\label{diffEuler}
\eear
plus the Poisson equation (\ref{poisson}).
Eliminating the magnetic force between these two equations and integrating, we arrive at the Bernoulli-type law,
\be
\delta h + \delta \Phi -\xp \delta \beta  = \mbox{constant}
\label{bernoulli}
\ee 
From these equations we can draw some key conclusions. Firstly, the magnetic force forbids the
establishment of chemical equilibrium (excluding the rather special case of a force-free field).
This point was also discussed at the end of Section~\ref{sec:formalism}.
Secondly, we can always write the ratio ${\bf F}_\rL /\rho_\rp$ as a perfect gradient (the same is true for 
$ {\bf F}_\rL/\rho$ only in the special case of a uniform proton fraction). 
This second property suggests that, within the present multifluid model, the calculation of the hydromagnetic
equilibrium should be based on the difference Euler equation (\ref{diffEuler}) rather than the total Euler equation
(\ref{avEuler}). The strategy becomes obvious by taking the curl of (\ref{diffEuler})
\be
{\bf \nabla} \times \left  \{ \frac{1}{\rho_\rp} {\bf F}_\rL \right   \} = 0 
\label{curlfree}
\ee
This is identical to the equation governing the hydromagnetic equilibrium in a single-fluid barotropic star
after the replacement $\rho \to \rho_\rp$. It should be also emphasized that the property (\ref{curlfree}) is characteristic
of multifluid systems; it is not obeyed by stratified systems described by single-fluid magnetohydrodynamics. This fact hints
at the possibility that hydromagnetic equilibrium in a stratified multifluid system may {\em differ} from that expected in
single-fluid systems \citep{reisenegger09}.

Hence, the calculation of the multifluid equilibrium should consist of the same steps as the ones taken in the more familiar barotropic problem: 
(i) the ${\bf B}$ field is determined by (\ref{curlfree}) and the vanishing azimuthal component $F^\varphi_\rL = 0$ 
(a consequence of axisymmetry) for a given background density profile (ii) eqns.~(\ref{diffEuler}), (\ref{bernoulli}) and 
(\ref{poisson}) can be used to find the remaining unknown functions $\delta \beta, \delta h, \delta \Phi$.
In this work we will concern ourselves with the first step since our main interest is the magnetic field configuration in equilibrium.
The second step would have been essential if we wished to calculate the deformation of the star's shape
due to the magnetic field (as in \citet{haskell08}).

The first step in our analysis is to decompose the magnetic field into poloidal and toroidal components,
\be
{\bf B} = {\bf B}_\rP + {\bf B}_\rT, \qquad {\bf B}_\rP = {\bf \nabla} S \times {\bf \nabla} \varphi, \qquad
{\bf B}_\rT = T {\bf \nabla}\varphi
\label{poltor}
\ee
where $S(r,\theta)$ and $T(r,\theta)$ are axisymmetric ``stream functions''  with ${\bf B} \cdot {\bf \nabla} S =0$
(we hereafter adopt standard spherical coordinates $r,\theta,\varphi$).
The explicit form of the magnetic field components is 
\be
B^r = \frac{\partial_\theta S}{r^2\sin\theta}, \qquad B^\theta = - \frac{\partial_r S}{r\sin\theta}, \qquad
B^\varphi = \frac{T}{r\sin\theta}
\label{compons}
\ee
It is easy to see that the magnetic field, when written in the form (\ref{poltor}), is automatically 
divergence-free.

As already mentioned, the 
assumption of axisymmetry requires that $F^\varphi_\rL =0 $. This means that we obtain
\be
{\bf B}_\rP \cdot {\bf \nabla} (\varpi B_\rT ) =0 \quad \to \quad {\bf \nabla} S \times {\bf \nabla} T =0
\label{T_S}
\ee
which shows that the two stream functions share the same level surfaces, i.e. we can write $T=T(S)$ 
(this is of course a classic result, see \citet{chandra56}).

In terms of the stream functions and after some straightforward manipulations the Lorentz force becomes
\be
{\bf F}_\rL = -\frac{1}{4\pi \varpi^2} \left [ \Delta_* S + T \frac{dT}{dS} \right ]\, {\bf \nabla} S
\equiv \cA {\bf \nabla} S
\label{forceS}
\ee
where $\varpi = r\sin\theta$ is the usual cylindrical radius and
\be
\Delta_* = \varpi^2 {\bf \nabla} \cdot \left ( \varpi^{-2} {\bf \nabla} \right )
\ee
is the so-called Grad-Shafranov operator. We also note that eqn.~(\ref{forceS}) defines the 
parameter $\cA$.

Specialising to spherical coordinates we can write
\be
\Delta_* S = \nabla^2 S - \frac{2}{r} \left \{ \, {\bf \hat{r}} \cdot {\bf \nabla} S
+ \cot\theta ( {\bf \hat{\theta}} \cdot {\bf \nabla} S ) \, \right \}
\label{opera}
\ee
It is also worth pointing out the relation between this operator and Chandrasekhar's ``$\Delta_5$'' operator
(e.g. \citet{chandra56}).
Momentarily switching to cylindrical coordinates $\varpi,z,\varphi$ we can easily show that
\be
\Delta_* S = \varpi^2 \left ( \partial^2_\varpi P + \frac{3}{\varpi} \partial_\varpi P + \partial^2_z P \right )
= \varpi^2 \Delta_5 P
\ee
where $P= S/\varpi^2$.

Inserting the Lorentz force (\ref{forceS}) in (\ref{curlfree}) we arrive at
\be
{\bf \nabla} \left ( \frac{\cA}{\rho_\rp} \right ) \times {\bf \nabla} S = 0
\label{GS2}
\ee
This is solved by (essentially a Grad-Shafranov equation) 
\be
\frac{\cA}{\rho_\rp} = F(S) 
\label{sol1}
\ee
where $F$ is an arbitrary function, which is in principle `user-specified'. 
However, the perturbative approach followed here (with the magnetic field treated as a perturbation 
on a non-magnetic background and $S \sim {\cal O}(B)$) dictates that the {\em unique} choice for $F(S)$ is
(see also \citet{ciolfi09})
\be
\frac{\cA}{\rho_\rp} = F(S) = c_0 + c_1 S
\label{sol2}
\ee
with $c_0,c_1$ constants (note that $c_0$ is allowed to be $\sim {\cal O}(S)$). 
Thus, the combination of (\ref{forceS}), (\ref{opera}) and (\ref{sol2}) leads to our `final' 
equation for the hydromagnetic equilibrium 
\be
\Delta_* S + T \frac{dT}{dS} = -4\pi \xp \rho r^2 \sin^2\theta \left (  c_0 + c_1 S \right )
\label{masterpde}
\ee

So far we have only discussed the equations pertaining to the stellar interior. For the exterior (the magnetosphere) 
we use the commonly adopted assumptions of a perfect vacuum and an irrotational field ${\bf\nabla} \times {\bf B}_{\rm ex} = 0$ 
(the index `${\rm ex}$' denotes an exterior quantity). Combined with axisymmetry, the irrotationality condition 
implies a purely poloidal, force-free exterior field. From (\ref{poltor}) and (\ref{forceS}) this means that 
the exterior magnetic field is described by
\be
T_{\rm ex}=0, \qquad \quad \Delta_* S_{\rm ex} =0
\ee


\section{Decomposition in multipoles}

The previous equations can be solved by means of an expansion in angular spherical harmonics (see, for example, \citet{ciolfi09}).
Before doing this, we note that  it is more convenient to work with the auxiliary function $\tilde{S}$ defined by
\be
S = \sin\theta \partial_\theta \tilde{S}
\ee
Decomposing in terms of standard spherical harmonics $Y^m_\ell$ (and fixing $m=0$ for our axisymmetric system),
 \be
\tilde{S} = \sum_{\ell \geq 1} a_\ell (r) Y^0_\ell (\theta)
\label{newS}
\ee
we find that $\Delta_* S$ takes the particularly simple form
\be
\Delta_* S = -\sum_{\ell \geq 1} \left \{ a_\ell^{\prime\prime} -\ell(\ell+1) \frac{a_\ell}{r^2}   \right \} \sin\theta \partial_\theta Y^0_\ell 
\label{Adeco}
\ee
where a prime denotes a radial derivative. The expansion (\ref{newS}) also makes direct contact with the multipolar 
structure of the ${\bf B}$ field itself. For example, we have
\be
B^r(r,\theta) = -\sum_{\ell \geq 1} \frac{\ell(\ell+1)}{r^2} a_\ell(r) Y^0_\ell (\theta) 
\ee


\subsection{Poloidal equilibrium}

In this Section we consider a purely {\em poloidal} magnetic field ($T=0$). After some straightforward algebra, 
we find that (\ref{masterpde}) and (\ref{Adeco}) lead to the recurrence relation 
(with $\ell \geq 0$)
\begin{multline}
(\ell-1) Q_\ell C_{\ell-1} - (\ell+2) Q_{\ell+1} C_{\ell+1} = 
\\
-4\pi r^2 \rho_0 \Bigg [ c_0 \left \{ Q_2 \delta^2_\ell -2Q_1 \delta^0_\ell 
\right \} + c_1 \Bigg \{\,  Q_\ell K_{\ell-1}\, \alpha_{\ell-1} 
\\
- Q_{\ell+1} N_{\ell+1} \,\alpha_{\ell+1} -(\ell-3) Q_\ell Q_{\ell-1}
Q_{\ell-2} \, \alpha_{\ell-3} 
\\
+ (\ell+4) Q_{\ell+1} Q_{\ell+2} Q_{\ell+3}\, \alpha_{\ell+3} \, \Bigg \}   \Bigg ]
\label{master}
\end{multline}
where
\bear
&& C_\ell = \alpha_\ell^{\prime\prime} - \frac{\ell (\ell+1)}{r^2} \alpha_\ell, \qquad
 Q^2_\ell = \frac{\ell^2}{4\ell^2-1}
\\
\nonumber \\
&& K_\ell = \ell \left (1 - Q^2_{\ell+1} - Q^2_{\ell+2} \right ) + (\ell+1) Q^2_\ell
\\
\nonumber \\
&& N_\ell = (\ell+1) \left ( 1 -Q^2_\ell -Q^2_{\ell-1} \right ) + \ell Q^2_{\ell +1}
\eear
and $\delta^n_\ell$ is the usual Kronecker-delta. A closer inspection of (\ref{master}) reveals that the even and odd 
$\ell$-multipoles decouple from each other.

In the exterior space the situation is much simpler, with all multipoles decoupling, 
\be
r^2 \frac{d^2  a_{\ell}^{\rm ex}}{dr^2} -\ell(\ell+1) a_\ell^{\rm ex} = 0 \quad
\to \quad a_\ell^{\rm ex} (r)  = \frac{d_\ell}{r^{\ell}}
\label{ext}
\ee
where $d_\ell$ is a constant.

The above equations are supplemented by boundary conditions at the stellar center and surface. 
Assuming a power-law behaviour near $r=0$  we find that the desired solution is
\be
\alpha_\ell \sim r^{\ell+1},  \qquad r \to 0
\ee   
At the surface the field can be {\em smoothly} matched to the vacuum exterior provided
both functions $\alpha_\ell, \alpha^\prime_\ell$ are continuous. Using (\ref{ext}) this requirement leads to
the surface condition
\be
\alpha_\ell^\prime = -\frac{\ell}{R}\, \alpha_\ell, \qquad r = R
\ee

We now consider the simplest possible configuration, namely, a purely {\em dipolar} field ($\alpha_1 \neq 0, \alpha_{\ell \geq 2} = 0 $), 
in which case the only non-trivial equations (\ref{master}) are:
\bear
&&  C_1 + 2\pi r^2 \rho_\rp \left ( 2 c_0 + \frac{8}{5} c_1 \alpha_1 \right ) = 0   \qquad (\ell=0)
\label{l0eqn}
\\
&& C_1 + 4\pi r^2 \rho_\rp \left ( c_0 + \frac{8}{7} c_1 \alpha  \right ) = 0    \qquad (\ell = 2)
\eear
These are mutually consistent provided $c_1 = 0$. Thus, the dipole poloidal field is described by a single
differential equation
\be
\alpha_1^{\prime\prime} - \frac{2}{r^2} \alpha_1 = -4\pi c_0 r^2 \xp \rho
\label{ode1}
\ee 
together with
\be
S(r,\theta) = -\lambda \alpha_1(r) \sin^2\theta, \qquad \lambda = \sqrt{\frac{3}{4\pi}} 
\ee


\subsection{Adding a toroidal component}
\label{sec:toro}

We now extend the analysis by adding a toroidal magnetic field component.
Unlike $F(S)$, the function $T(S)$ cannot be uniquely specified within the perturbation scheme.

A general form for $T(S)$ that complies with the spirit of our perturbation scheme is (also used by \citet{ciolfi09})
\be
T(S) = -\zeta_0 S \left (\, \left |\frac{S}{S_0} \right | -1 \, \right ) 
\Theta \left ( \left |\frac{S}{S_0} \right | -1 \right )
\label{Tform}
\ee
where $\zeta_0$ and $S_0$ are constant parameters. The former parameter represents the overall toroidal
field strength while the latter designates the poloidal stream function surface which acts as the boundary for
the toroidal field (this is the role of the step-flunction, $\Theta$, in (\ref{Tform})). 
After some simple manipulations (\ref{Tform}) leads to
\be
T \frac{dT}{dS} = \zeta_0^2  S \left (\, 1 -3  \frac{|S|}{|S_0|} + 2  \frac{S^2}{S_0^2} \, \right )
 \Theta \left ( \left |\frac{S}{S_0} \right | -1 \right )
\ee

A common choice in this problem area, and the one we will adopt for the rest of this paper, 
is to choose $S_0$ as the {\em outermost closed} stream surface that is tangent to the surface of the star. 
As a result of this choice the $T$ function (\ref{Tform}) ensures that the toroidal field is vanishing at the 
surface (as required by the boundary conditions). Moreover, this specific choice for $T(S)$ 
has the advantage of ``placing'' the toroidal component in the vicinity of the poloidal field's 
`neutral line' (i.e. the locus of points where $B_\rP = 0$) ;
this configuration is likely to be dynamically stable as suggested by the work of \citet{tayler73}.

We again assume a purely dipolar field which obviously now has a mixed poloidal-toroidal character. This time, however, we find 
that the general recurrence relation for the $\alpha_\ell$  coefficients does not terminate at the first equation
(as was the case for the purely poloidal field). In other words within the present framework of a mixed magnetic field we 
cannot have a purely dipolar geometry.

We sidestep this difficulty by nevertheless assuming a dipolar field, truncating the recurrence 
relation at the first equation and setting $c_1=0$. This simplification is not as radical as it may seem.
According to the results of \citet{ciolfi09} 
we would expect the addition of higher odd-order multipoles $\ell=3,5$ etcetera to have only a minor impact on the equilibrium structure
of the $\ell=1$ configuration. The veracity of this assertion is further supported by the close agreement between the results of
this paper and those obtained in the companion paper \citep{LAG} which considers a generic magnetic field geometry. 


\subsection{Numerical solution}

Following the approach outlined in the preceding paragraph we find that (\ref{ode1}) is replaced 
by
\be
\alpha_1^{\prime\prime} - \frac{2}{r^2} \alpha_1 = -4\pi c_0 r^2 \xp \rho + \frac{3}{4} \zeta^2_0 \cI
\label{ode2}
\ee
where the toroidal term contains the angular integral
\begin{multline}
\cI(r) = \int^\pi_0 d\theta \sin\theta
\left (\, S -3S  \frac{|S|}{|S_0|} + \frac{2S^3}{S_0^2} \, \right )
 \Theta \left ( \left |\frac{S}{S_0} \right | -1 \right ) \\
=  -\lambda \alpha_1\int^\pi_0 d\theta \sin^3\theta   \Bigg \{ 1 -3 \left | \frac{\lambda \alpha_1}{S_0}   
\right | \sin^2 \theta \\
+ 2 \left ( \frac{\lambda \alpha_1}{S_0} \right )^2 \sin^4 \theta \Bigg \}
\Theta \left ( \left |\frac{\lambda \alpha_1}{S_0} \right | 
\sin^2\theta -1 \right )
\label{angular}
\end{multline}
To facilitate the numerical integration of (\ref{ode2}) (or (\ref{ode1})) we introduce a new set of dimensionless parameters.
First, we define $x = r/R$ which also allows us to write
\be
\rho  = \frac{M}{R^3} f(x) 
\ee
where $M$ and $R$ are the stellar mass and radius. 
Expressed in terms of the radial field at the magnetic pole we have
\be
\alpha_1 (R) = -\frac{B_p R^2}{2 \lambda}, \qquad B_p = B^r (r=R,\theta=0)
\ee
We can then normalise $\alpha_1$ as
\be
\tilde{\alpha}_1 = \alpha_1 \frac{2 \lambda}{B_p R^2}
\ee
In terms of the new variables (\ref{ode2}) becomes
\be
\frac{d^2 \tilde{\alpha}_1 }{dx^2}  - \frac{2}{x^2} \tilde{\alpha}_1 = -4\pi d_0\, x^2 \xp f(x) 
+ \frac{3}{4} \tilde{\zeta}_0^2\, \tilde{\cI}(x)
\label{ode3}
\ee
where
\be
\tilde{\cI}  = \frac{2\lambda}{B_p R^2}\, \cI, \quad  
d_0 = \frac{2\lambda R^2}{B_p} c_0 \rho_0, \quad \tilde{\zeta}_0 = \zeta_0 R 
\ee
are all dimensionless parameters. The boundary conditions at the
stellar center and surface accordingly change to
\be
\tilde{\alpha}_1 (x\to0) \sim x^2, \qquad  \frac{d \tilde{\alpha}_1}{dx} (1) = -\tilde{\alpha}_1 (1)
\label{newbcs}
\ee

Our next task is to simplify the toroidal term $\tilde{\cI}(x)$. The angular integrals in  (\ref{angular})  
are non-vanishing within the latitudinal interval $ \theta_0(r) < \theta < \pi-\theta_0(r) $ with $\theta_0$ defined 
as
\be 
| \kappa \tilde{\alpha}_1 | \sin^2\theta_0 = 1, \qquad \kappa = B_p R^2 / 2 S_0 = \mbox{const.}
\ee
We can then carry out the angular integrations,
\begin{multline}
\tilde{\cI}(x) = -2 \lambda \tilde{\alpha}_1 \cos\theta_0 \Bigg \{\, \frac{1}{3} \left (2 + \sin^2\theta_0 \right )   
\\
-\frac{1}{5}  \left | \kappa \tilde{\alpha}_1 \right | \left ( 3\sin^4 \theta_0 + 4 \sin^2\theta_0 + 8 \right )
\\ 
+ \frac{2}{35} \left ( \kappa \tilde{\alpha}_1 \right )^2 \left ( 5 \sin^6\theta_0 + 6\sin^4\theta_0 
+8\sin^2\theta_0 + 16  \right ) \, \Bigg \}
\end{multline}
Finally, given the previous normalisations, we define the dimensionless magnetic field components 
\bear
&& b^r = \frac{B^r}{B_p} =  - \frac{\tilde{\alpha}_1}{x^2} \cos\theta
\\
\nonumber \\
&& b^\theta = \frac{B^\theta}{B_p} = \frac{1}{2x} \frac{d \tilde{\alpha}_1}{dx} \sin\theta
\\
\nonumber \\
&& b^\varphi = \frac{B^\varphi}{B_p} =\frac{\tilde{\zeta}_0 \tilde{\alpha}_1}{2x} \sin\theta
\left (\, \left |\kappa \tilde{\alpha}_1 \right | \sin^2\theta -1 \, \right ) 
\Theta \left ( \left |\kappa \tilde{\alpha}_1 \right | \sin^2\theta -1 \right )
\nonumber \\
\eear
noting again that $B_p$ is the polar magnetic field strength.


\section{Results: multifluid hydromagnetic equilibrium}

\begin{figure}
\centerline{\includegraphics[height=4cm,clip]{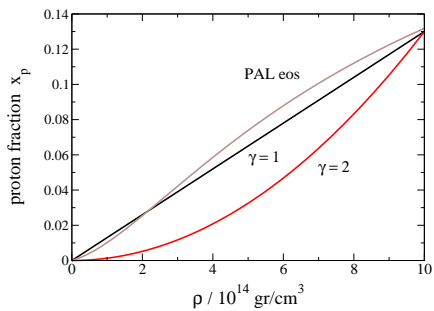}}
\caption{The proton fraction profile $\xp (\rho)$, eqn.~(\ref{xp}), for $\gamma=1$ and $\gamma=2$ is compared to the \citet{kaminker01} fit to 
a typical PAL equation of state \citep{pal}. The central density is taken to be $\rho_c = 10^{15}\,\mbox{gr}/\mbox{cm}^3$. This example shows 
that it is reasonable to model the core proton fraction as linear.}
\label{fig:xp_profile}
\end{figure}

In this Section we construct hydromagnetic equilibria by solving eqn.~(\ref{ode3}) for the radial function $\tilde{\alpha}_1$.
For the non-magnetic background we choose a stellar model with a $n=1$ polytropic density profile
\be
\rho = \frac{M}{4 R^3x} \sin ( \pi x) \quad \to \quad f(x) = \frac{\sin(\pi x)}{4x}
\ee
Given the overall precision of our modelling this choice is a good 
approximation to more realistic equations of state. For the proton fraction we use 
\be
x_\rp = 0.13 \left(\frac{\rho}{\rho_c}\right )^\gamma
\label{xp}
\ee
where $\rho_c = \pi M/4R^3$ is the central density and $\gamma$ a constant. This expression is motivated by the work of \citet{RG92} 
who determined $\xp$ for a Fermi mixture of non-interacting neutrons, protons and electrons. Comparing the $\gamma=1$ form of (\ref{xp}) 
against the profile of a typical representative of realistic equations of state we find good agreement, see Fig.~\ref{fig:xp_profile}.
It should be also pointed out that the prescription (\ref{xp}) is valid only in the stellar core and not in the region of the crust 
or the surface. We nevertheless use it for the entire stellar volume; this should be a reasonably accurate approximation 
(see discussion at the end of Section~\ref{sec:current}).

For the purpose of this paper we have introduced the phenomenological parameter $\gamma$ which controls the relative proton fraction to density 
scale-heights,
\be
L_\rx = \frac{\xp}{|\nabla \xp|} =  \frac{1}{\gamma} \frac{\rho}{\rho^\prime}
\label{height}
\ee 
and as such it is an effective measure of the stratification. Although a  value of $\gamma$ significantly different from
1 may not be such a good approximation to realistic $\xp$ profiles (see Fig.~\ref{fig:xp_profile}), 
it is nevertheless of interest to experiment with this parameter to gain intuition about the effect of stratification on hydromagnetic equilibrium.

The numerical integration of (\ref{ode3}) with the boundary conditions (\ref{newbcs}) is straightforward. 
For each calculated hydromagnetic equilibrium we need to specify three parameters: the proton fraction power-law $\gamma$ 
in (\ref{xp}), the toroidal amplitude $\tilde{\zeta}_0$ and the ratio $\kappa = B_p R^2/2S_0 $ (without loss of generality we can 
set $\kappa=1$). The remaining constant $d_0$ is fixed during the integration itself.

A sample of hydromagnetic equilibria is shown in Figures~\ref{fig:Bprofile} and \ref{fig:streams}.
Let us first discuss Fig.~\ref{fig:Bprofile} which displays the radial profile of the normalised magnetic field 
${\bf b}(x,\theta)$. In particular we show  $b^r$ along the polar direction $\theta=0$ and  $b^\theta, b^\varphi$ in the
equatorial plane $\theta=\pi/2$. We have used $\tilde{\zeta}_0 = 0, 10, 20$ (the first is just a poloidal field) 
and for each case we have chosen three different stellar models: $\gamma=1$, which is a model with 
`canonical' composition stratification, $\gamma=2$, which represents a fiducial strongly-stratified model 
and finally a barotropic model, which formally corresponds to $\xp=1$. 

The equilibria in Fig.~\ref{fig:Bprofile}  are also shown in Fig.~\ref{fig:streams} in the form of contour plots  
of the normalised stream function $S/S_0$, projected on an arbitrary meridional plane. The two figures have been arranged such that each 
panel is in  one to one correspondence. 
 
The nine displayed equilibria share some common properties. They all feature a single neutral line at 
the equatorial plane (where $B^\theta (x_n,\pi/2) = 0$); when a toroidal field is present they all resemble a ``twisted-torus''
configuration with the toroidal field roughly occupying the ``hole'' in the vicinity of the neutral line.
From this point of view, the structure of our multifluid equilibria is  similar to the existing barotropic ones 
\citep{ciolfi09,lander09,ciolfi10} as well as the numerical equilibria of \citet{braith06,braith09} (the latter ones
represent stable hydromagnetic equilibria of a stellar model obeying an ideal gas equation of state -- see discussion
at the end of Section~\ref{sec:strato}).


\subsection{The role of stratification}
\label{sec:strato}

Is composition stratification (non-uniform $\xp$) an important factor in the hydromagnetic equilibrium 
of multifluid neutron stars? 

\begin{figure}
\centerline{\includegraphics[height=7cm,clip]{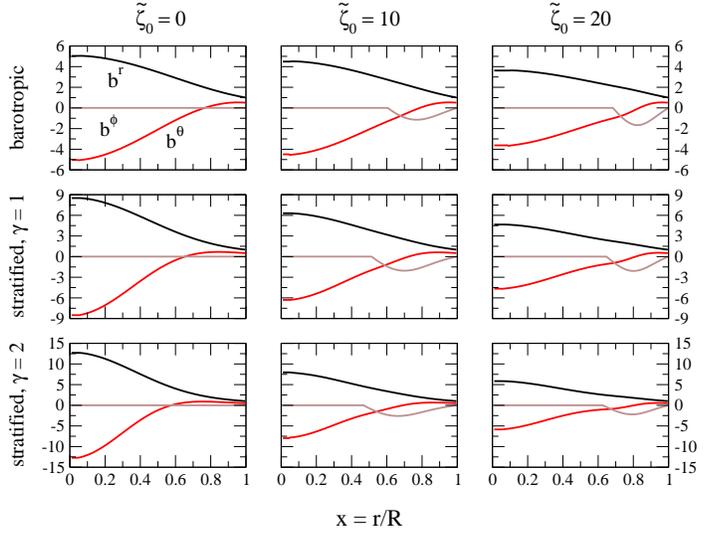}}
\caption{The radial profile of the normalised magnetic field components $b^i = B^i/B_p$ 
($B_p$ is the field at the magnetic pole and $i=\{r,\theta,\varphi\}$) for three stellar models:
(i) a barotrope (top row), (ii) a stratified model with a $\gamma=1$ profile for the proton fraction $\xp$ (middle row) and (iii) 
a stratified model with $\gamma=2$ (bottom row). Each column corresponds to a different choice for the toroidal field
amplitude, $\tilde{\zeta}_0 = 0,10,20$. The $b^r$ field is calculated along the $\theta =0$ direction
while the $b^\theta$ and $b^\varphi$ components are calculated along $\theta=\pi/2$. Note that the numerical y-axis scale  
is kept fixed between the columns of a given row.}
\label{fig:Bprofile}
\end{figure}

\begin{figure}
\centerline{\includegraphics[height=4cm,clip]{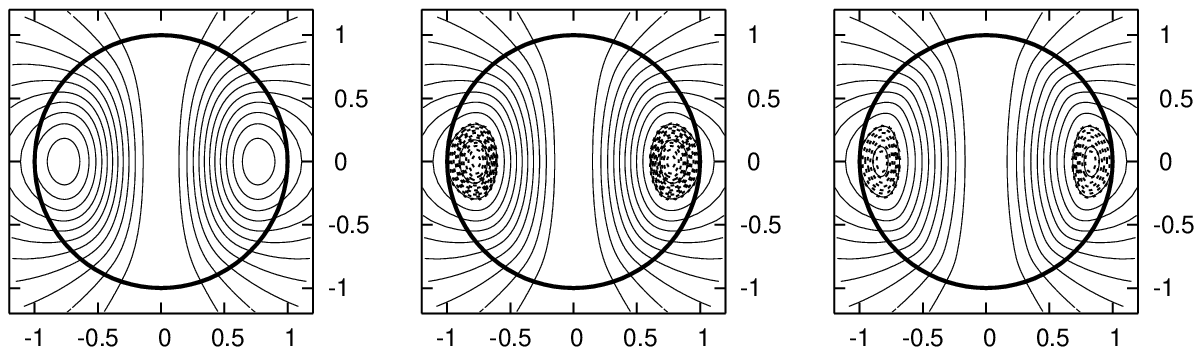}}
\centerline{\includegraphics[height=4cm,clip]{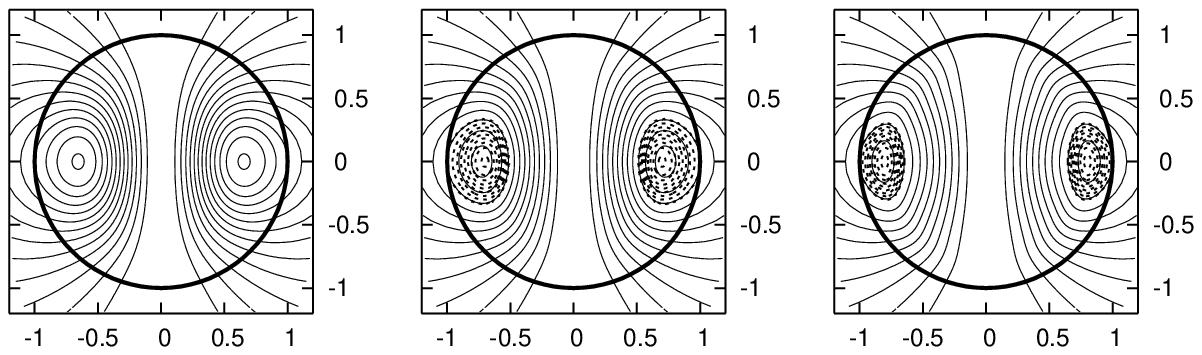}}
\centerline{\includegraphics[height=4cm,clip]{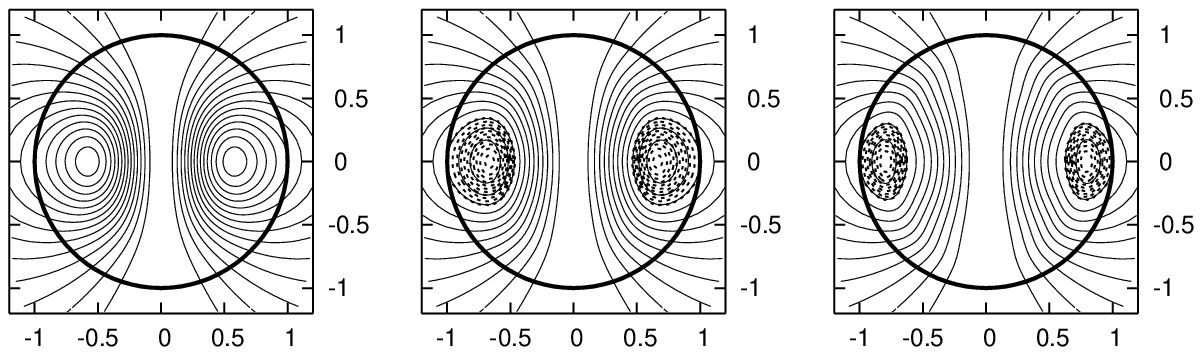}}
\caption{The hydromagnetic equilibria of Fig.~\ref{fig:Bprofile} shown as contour plots of the normalised stream
function $S/S_0$. Each panel corresponds to a panel of Fig.~\ref{fig:Bprofile} located in the same row and column.
Top row: barotropic models. Middle row: stratified $\gamma=1$ models. Bottom row: stratified $\gamma=2$ models.
Columns from left to right: $\tilde{\zeta}_0 =0, 10, 20$. The thick circular line represents the stellar surface.
The toroidal component $b^\varphi$ occupies the region bounded by the outermost closed poloidal streamline and it is shown 
with dashed lines.}
\label{fig:streams}
\end{figure}

Based on the evidence provided by  the poloidal equilibria in Figs.~\ref{fig:Bprofile} and \ref{fig:streams}, 
the answer is clearly yes. A stronger stratification (i.e. a larger $\gamma$) pushes the 
location of the neutral line {\em inwards}. The most extreme example shown here (the $\gamma=2$ equilibrium) has its neutral line at 
$x_n \approx 0.6$ as compared to the $x_n \approx 0.8$ of the barotropic equilibrium, c.f. the left column panels in Figs.~\ref{fig:Bprofile} 
and \ref{fig:streams}. Intuitively, this behaviour makes sense since the force term in (\ref{avEuler}) due to the stratification points inwards
(the stratification is stable).

Another interesting effect of stratification regards the relative poloidal field amplitude between the stellar 
surface and center. This is again evident in Fig.~\ref{fig:Bprofile} by looking at each column separately (keeping in mind the changing
numerical scale). A stronger stratification implies a larger central poloidal field $B_\rP$ relative to its
surface value (which is the same for all equilibria shown here).

The addition of a toroidal component invariably moves the neutral line {\em outwards}. As a result of this, 
an increasing toroidal amplitude $\tilde{\zeta}_0$ results in a smaller portion of the stellar volume 
being occupied by the toroidal field. At the same time, the overall magnetic configuration appears to
take similar forms regardless of stratification. For instance, all $\tilde{\zeta}_0 = 20$ equilibria look
similar to the single-fluid barotropic equilibria of \citet{ciolfi09}.

Another way of looking at the impact of stratification on the magnetic field structure is by calculating the
magnetic energy stored in the poloidal and toroidal components. These are given by
\be
E_\rP = \frac{1}{8\pi} \int dV B^2_\rP, \qquad E_\rT = \frac{1}{8\pi} \int dV B^2_\rT 
\ee
where the integrals are taken over the stellar volume.

\begin{figure}
\centerline{\includegraphics[height=4.5cm,clip]{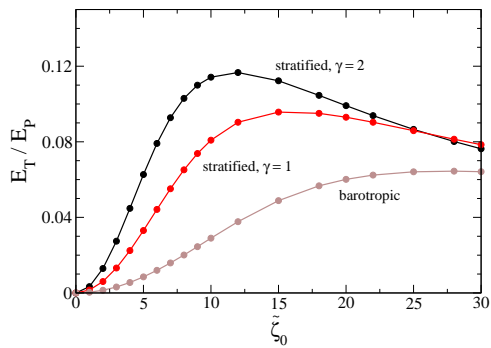}}
\caption{The ratio $E_\rT/E_\rP$ of toroidal to poloidal magnetic field energy as a function of the dimensionless toroidal amplitude
$\tilde{\zeta}_0$. The stellar models are the same as in the previous Figures, namely, a barotrope, and  $\gamma=1, ~\gamma=2$ proton fraction
stratification. For the models shown here the maximum toroidal field strength is $B_\varphi^{\rm max} \lesssim 4 B_p$, where $B_p$ is the
polar magnetic field.}
\label{fig:energs}
\end{figure}

A useful quantity to display is the ratio $E_\rT/E_\rP$; which can easily be expressed in terms of the normalised parameters 
of Section~\ref{sec:toro}. Our results are shown in Fig.~\ref{fig:energs}, where we have again considered the same three models 
as before (barotropic, $\gamma=1$, $\gamma=2$).

The message from Fig.~\ref{fig:energs} is rather clear: stratification leads to hydromagnetic equilibria with a {\em higher}
$E_\rT/E_\rP$ energy ratio content with respect to their barotropic counterparts. However, even for our most strongly stratified model 
the actual energy ratio ``saturates'' at a small value, never exceeding the level of  $E_\rT/E_\rP \sim 0.1$. 
This is a direct consequence of the outward  ``motion'' of the neutral line with an increasing toroidal component.

In terms of the actual strength of the toroidal field, we note (Fig.~\ref{fig:Bprofile}) that the maximum relative  magnitude
$|B^\varphi/B_p|$ (located at $\theta=\pi/2$ ) never exceeds a factor $\sim 3-5$. 

At this point it would be useful to confront the results in Figs.~\ref{fig:Bprofile} and \ref{fig:streams} with those by Braithwaite 
and collaborators \citep{braith06,braith09}. 
This comparison is of particular interest given that Braithwaite's stellar model is the only (previous) non-barotropic model in the literature. 
However, this is where the similarities between the two models end. Braithwaite utilises an ideal gas equation of state to build 
a stellar model characterised by a stratification in the temperature/entropy profile. This may be an appropriate model
for a main sequence star but it is not so good for neutron stars which are highly isothermal/isentropic objects with stratification
associated with  matter composition. Despite this fundamental difference, the stable quasi-axisymmetric twisted-torus equilibria obtained
by Braithwaite look roughly the same as the ones shown in our Fig.~\ref{fig:streams}. However, Braithwaite's equilibria have a much 
higher energy ratio  $E_\rT/E_\rP \gg 1 $; in that sense they are dominated by the toroidal field. 
In our opinion, the most natural explanation for this variance in the magnetic energies
$E_\rT/E_\rP$ appears to be the difference in the physics of stratification between the two models. 
An alternative (and less likely) explanation could be that the equilibria in Fig.~\ref{fig:streams} are for some reason 
dynamically unstable and there is a second, yet undiscovered, class of axisymmetric hydromagnetic solutions with much stronger toroidal fields.


\subsection{The effect of a surface current}
\label{sec:current}

A different aspect of hydromagnetic equilibrium in neutron stars that has received little attention so far
is the possibility of having electric currents located at the stellar surface. Realistic neutron stars could
in fact harbour such currents as a result of the presence of the magnetosphere.

A detailed modelling of surface currents in neutron stars is beyond the scope of this paper. Indeed, we are agnostic on the
question of how such currents could be generated and maintained. We may nevertheless
learn something useful by following a simple phenomenological approach where the presence of a surface current is
mimicked by a modification of the magnetic field at the stellar surface. A generic consequence of
a surface current is to make the magnetic field discontinuous at the surface (this is a direct consequence of Amp\`ere's law
$\nabla \times {\bf B} = 4\pi {\bf J}/c$). 
The idea then is to adopt a slightly different set of surface boundary conditions which incorporate such a discontinuity. 
A possible choice is\footnote{The radial component $B^r$ is always continuous at an interface as a result of 
${\bf \nabla} \cdot {\bf B} =0$.}
\be
b^r  = b^r_{\rm ex}, \qquad b^\theta = \xi b^\theta_{\rm ex}, \qquad  b^\varphi = 0  \qquad (x = 1)
\label{newbc}
\ee 
which corresponds to an azimuthal surface current. The parameter $\xi$ controls the surface current `strength' 
(with $\xi=1$ representing the previous case of a continuous ${\bf B}$ field and vanishing surface current). 
In terms of the $\tilde{\alpha}_1$ function the new set of surface conditions (\ref{newbc}) is equivalent to $d\tilde{\alpha_1}/dx = -\xi \tilde{\alpha}_1 $.

\begin{figure}
\centerline{\includegraphics[height=4cm,clip]{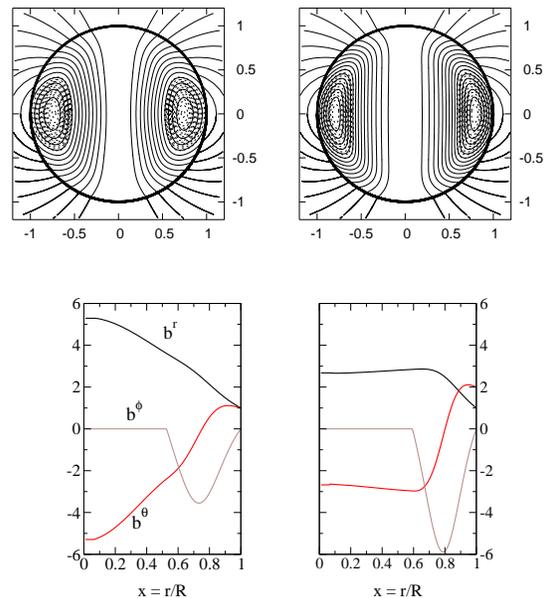}}
\centerline{\includegraphics[height=4cm,clip]{Bprofile_x1_z10_bc2bc4.eps}}
\caption{Hydromagnetic equilibria for a $\gamma=1, \tilde{\zeta}_0 = 10$ model with a ``surface current'' 
(see Section~\ref{sec:current} for details). Each contour plot corresponds to the radial profiles plot below it
(see Figs.~\ref{fig:Bprofile} and \ref{fig:streams} for a description). Left panels: $\xi = 2 $. Right panels: $\xi = 4$. 
The toroidal to poloidal energy ratio for these two equilibria is $E_\rT/E_\rP = 0.267$ (left), $E_\rT/E_\rP =  0.654$ (right).}
\label{fig:current}
\end{figure}

For the purpose of illustration we have constructed two equilibria using the stratified $\gamma=1$ model with $\tilde{\zeta}_0=10$
and for two fiducial values $\xi=2$ and $\xi=4$ for the discontinuity parameter. The results are shown in Fig.~\ref{fig:current}.
The most notable difference with respect to the previous equilibria relates to the larger toroidal field component and energy ratio $E_\rT/E_\rP$; 
the latter can easily exceed the few percent level. Indeed it is now possible to build equilibria with comparable energies, i.e. 
$E_\rT \sim E_\rP$ (e.g. the $\xi=4$ model in Fig.~\ref{fig:current}). In a sense, the equilibria discussed in this section
can be viewed as the middle ground between magnetic configurations that are fully confined within the star and can have
$E_\rT \gg E_\rP$ (e.g. \citet{haskell08}) and the configurations discussed in the previous sections which smoothly extend
to the exterior space and have $E_\rT \ll E_\rP$. If nothing else, this result highlights the need for
a detailed study of the effect that a non-vacuum magnetosphere could have on the structure of the hydromagnetic equilibrium 
in realistic neutron stars.

Before moving on, we note that the truncated recurrence relation also allow a second class of solutions,  corresponding to a different value 
of the constant $d_0$. This new family of solutions appears only for mixed poloidal-toroidal fields, in both barotropic and stratified models, 
but as it is likely to be an artifact of the truncation in the multipole expansion of the stream function $S$ (c.f. the discussion of 
\citet{colaiuda08} and \citet{ciolfi09}) we will not consider it in detail here.

We also carried out a set of computations in order to  assess the effect of adding a fiducial spherical layer of matter 
with a barotropic (i.e. $\xp=1$) equation of state. This construction can be considered as a toy model of a neutron star with a 
crust. In our model the `crust' was added to  the outer $x > 0.9$ region of the star. Not surprisingly, the resulting equilibria only 
display minor differences from our previous results. 

Finally, we carried out computations of equilibria with a non-zero constant $c_1$ (see eqns.~(\ref{masterpde}) and (\ref{l0eqn})). 
For a broad range of $c_1$ values (including the case where $c_0=0$ and $c_1$ is fixed during the integration itself) the resulting
solutions display only minor differences with respect to the ones shown in Figs.~\ref{fig:Bprofile} and \ref{fig:streams}.


\section{Concluding discussion}

This paper, together with its companion \citep{LAG}, represents the first detailed study of 
the hydromagnetic equilibrium in multifluid neutron stars. In spite of the obvious simplicity of our models 
(axisymmetry, dipolar magnetic field geometry, omission of proton 
superconductivity) we have managed to  probe important aspects of multifluid magnetohydrodynamics. 

We have shown that the problem of hydromagnetic equilibrium in a multifluid system, consisting of superfluid neutrons and normal protons,
can be formulated in terms of a Grad-Shafranov equation despite the inherent non-barotropicity of the system due to the 
varying composition.

The main result of this work, summarised in Figs.~\ref{fig:Bprofile}-\ref{fig:energs}, concerns the impact of 
the composition stratification (expressed in terms of the proton fraction gradient ${\bf \nabla} \xp$) on the structure of the magnetic
field in equilibrium. The obtained equilibria are of the twisted-torus type, containing a mixture of poloidal and toroidal
magnetic field components. 
We find that stratification generally increases the energy stored in the toroidal field by allowing
a larger portion of the stellar volume to be occupied by it. However, even for unrealistically strong stratification 
(our $\gamma=2$ model), we conclude that the produced equilibria are still dominated by the poloidal field, with the 
toroidal magnetic energy never exceeding a fraction  $\sim 0.1$ of the poloidal energy (Fig.~\ref{fig:energs}). 
At first glance, our results seem at odds with the non-barotropic twisted-torus equilibria of \citet{braith06,braith09}, in the sense 
that the latter are dominated by the toroidal field. As discussed in Section~\ref{sec:strato} the discrepancy may have a perfectly 
natural explanation in the different type of stratification assumed in the underlying stellar models. Nonetheless, this issue needs to be clarified
in future work.

The main conclusions of this paper are broadly supported by the more detailed modelling of \citet{LAG}, which extends the present work 
to general, non-dipolar magnetic configurations using a wider class of non-barotropic equations of state. The same work also lays the 
basis for the modelling of more realistic equilibria with type II superconductivity.

A crucial issue not addressed here is that of the {\em stability} of the calculated equilibria. Recent
numerical simulations \citep{braith08,lasky11,ciolfi11} have revealed the existence of dynamically stable {\em non-axisymmetric} equilibria
in both barotropic and non-barotropic (ideal gas) stellar models. Thus, it is of obvious importance to assess the stability of axisymmetric
equilibria like the ones obtained here.

Our multifluid hydromagnetic equilibria provide an improved understanding of different aspects of magnetar physics.
Taken at face value our results suggest that the surface magnetic field in magnetars (e.g. as inferred by the standard
spin-down formula \citep{STbook}) is also a good indicator of the interior magnetic field strength. More specifically, and contrary to
a widespread belief in the literature, the maximum toroidal field in the stellar interior is likely to be
{\em comparable} to the value at the magnetic pole. This result would suggest that the  reservoir of magnetic energy
available for powering giant flares in magnetars is constrained to a level indicated by the surface field. This may impact
on the viability of some proposed scenarios.

Finally, the moderate structural differences in the magnetic field
geometry caused by the stratification in a multifluid system is likely to have some impact on the ongoing theoretical effort
to interpret the quasi-periodic oscillations observed during giant flares as magnetar pulsation modes (e.g. \citet{levin11,gabler11,colaiuda11}). 
It should be clear that any model aiming to confront the real data must incorporate the key aspects of multifluid magnetohydrodynamics.


\section*{Acknowledgments}

KG is supported by an Alexander von Humboldt fellowship and by the German Science Foundation (DFG) via SFB/TR7.
NA acknowledges support from STFC in the UK through grant number PP/E001025/1. 
SKL acknowledges funding from the European Science Foundation (ESF) for the activity entitled `The New Physics of
Compact Stars'.


\end{document}